\documentclass[preprint2,twocolumn,times,tighten]{aastex63}

\usepackage{graphicx,times}
\usepackage{subfigure}
\newcommand{\be}{\begin{equation}}
\usepackage{threeparttable}
\usepackage{booktabs}
\newcommand{\ee}{\end{equation}}
\newcommand{\bea}{\begin{eqnarray}}
\newcommand{\eea}{\end{eqnarray}}

\usepackage{float}
\usepackage{amsmath}
\usepackage{cases}
\usepackage{longtable}
\usepackage{hyperref}
\usepackage{epstopdf}
\usepackage{amsmath,bm}
\usepackage{amssymb}
\usepackage{natbib}
\usepackage{morefloats}
\usepackage{multirow}
\usepackage{array}
\usepackage{verbatim}
\usepackage{color}
\usepackage{CJK}
\usepackage{lineno}
\begin{document}
\begin{CJK*}{UTF8}{gbsn}

\title{Numerical testing of mirror diffusion of cosmic rays}

\email{chao.zhang@ufl.edu;
xusiyao@ufl.edu}

\author[0009-0001-4012-2892]{Chao Zhang}
\affiliation{Department of Physics, University of Florida, 2001 Museum Rd., Gainesville, FL 32611, USA}

\author[0000-0002-0458-7828]{Siyao Xu\footnote{NASA Hubble Fellow}}
\affiliation{Department of Physics, University of Florida, 2001 Museum Rd., Gainesville, FL 32611, USA}

\begin{abstract}

The tension between recent observations and theories on cosmic ray (CR) diffusion necessitates exploration of new CR diffusion mechanisms. We perform the first numerical study on the mirror diffusion of CRs that is recently proposed by \cite{lazarian2021mirroring}. We demonstrate that the perpendicular superdiffusion of turbulent magnetic fields and magnetic mirroring that naturally arise in magnetohydrodynamic (MHD) turbulence are the two essential physical ingredients for the mirror diffusion to happen. In supersonic, subsonic, and incompressible MHD turbulence, with the pitch angles of CRs repeatedly crossing $90^\circ$ due to the mirror reflection, we find that the mirror diffusion strongly enhances the confinement of CRs, and their pitch-angle-dependent parallel mean free path can be much smaller than the injection scale of turbulence. With the stochastic change of pitch angles due to gyroresonant scattering, CRs stochastically undergo slow mirror diffusion at relatively large pitch angles and fast scattering diffusion at smaller pitch angles, resulting in a L\'{e}vy-flight-like propagation.

\end{abstract}

\section{Introduction}

Cosmic ray (CR) diffusion in turbulent and magnetized media is a fundamental problem in space physics and astrophysics. Theoretical studies of CR diffusion have improved our understanding of important astrophysical processes, e.g., planetary and star formation (\cite{rodgers2020astro4}; \cite{padovani2020astro41}; \cite{semenov2021cosmic}), evolution of galaxies and galaxy clusters (e.g.,\cite{guo2008feedback}; \cite{brunetti2014cosmic}; \cite{holguin2019role}; \cite{hopkins2020but}; \cite{quataert2022physics}). Recently, the new-generation CR observations challenge the existing theoretical paradigm of CR diffusion (e.g., \cite{gabici2019origin}; \cite{evoli2020ams}; \cite{amato2021particle}; \cite{fornieri2021theory}), which urges us to reexamine the fundamental physics of CR diffusion. 

CR diffusion strongly depends on the properties of magnetohydrodynamic (MHD) turbulence that they interact with. Direct simulations of MHD turbulence enable numerical testing of  the theoretical models of MHD turbulence (e.g., \cite{maron2001simulations}; \cite{cho2000anisotropy}; \cite{cho2002compressible}; \cite{cho2003compressible}) and CR diffusion (e.g., \cite{beresnyak2011numerical}; \cite{xuyan2013}; \cite{cohet2016cosmic}; \cite{yue2022superdiffusion};  \cite{sampson2023turbulent}; \cite{gao2023diffusion}). In the presence of magnetic fields, the CR diffusion perpendicular and parallel to the magnetic field needs to be treated separately.
Within the inertial range of MHD turbulence, the superdiffusion of turbulent magnetic fields (\cite{LV99reconnection}) causes the perpendicular superdiffusion of CRs (\cite{xuyan2013}; \cite{lazarian2014superdiffusion}; \cite{yue2022superdiffusion}).

In the direction parallel to the magnetic field, most earlier studies are focused on the diffusion induced by gyroresonant scattering. Due to the scale-dependent anisotropy of Alfv\'en and slow modes in MHD turbulence (\cite{gs95}; \cite{LV99reconnection}), Alfv\'en and slow modes are inefficient in scattering the CRs with gyroradii much smaller than the turbulence injection scale (\cite{chandran2000mirrorconfinement}; \cite{yan2002scattering}; \cite{xu2020trapping}), while fast modes are identified as the more efficient agent of scattering (\cite{yan2004cosmic}). More recently, strong scattering of CRs by sharp intermittent magnetic field bends in MHD turbulence is proposed to be important for affecting CR parallel diffusion (\cite{lemoine2023pitch}; \cite{kempski2023cosmic}; \cite{butsky2023constraining}). The enhanced local scattering can be associated with the intermittent and fractal structure of MHD turbulence (\cite{isliker2003random}). 

\begin{table*}[htb!]
\centering
\caption{Parameters of MHD turbulence simulations\label{data} }
\centering    
\begin{tabular}{|l||l l l l l l|} 
 \hline
   MHD Turbulence & $M_A$ & $M_s$ & $\beta$ & Resolution & $L_\text{inj}$ & $l_\text{diss}$ \\
 \hline\hline
  Compressible subsonic (M1) & 0.68 & 0.60 & 2.569 & $792^3$ & $\approx 400$ & $\approx$ 25\\
 \hline
 Compressible supersonic (M2)   & 0.58 & 10.50 & 0.006 & $512^3$ & $\approx 250$ & $\approx$ 20\\
  \hline
 Incompressible   (M3) & 0.80 & \textendash & \textendash & $512^3$ & $\approx 200$ & $\approx$ 10  \\
 \hline
\end{tabular}
\end{table*}

The resonant scattering faces its long-standing $90^\circ$ problem in the framework of the quasi-linear theory (\cite{90degreejokipii1966}). To resolve this issue, nonresonant interactions such as magnetic mirroring was explored (\cite{fermi1949origin}; \cite{cesarsky1973mirror}, CK73 henceforth; \cite{noerdlinger1968improved}; \cite{klepach1995propagation}; \cite{chandran2000mirrorconfinement}). This consideration of the mirroring effect naturally solves the $90^\circ$ problem and limits the pitch angle range for gyroresonant scattering (CK73; \cite{xu2020trapping}). Magnetic mirrors can naturally form in MHD turbulence due to compressions of magnetic fields, which are induced by pseudo-Alfv\'enic modes in an incompressible medium and slow and fast modes in a compressible medium. Unlike the trapping effect of magnetic mirrors considered for compressible MHD waves in CK73, \cite{lazarian2021mirroring} (LX21 henceforth) found that in MHD turbulence due to the perpendicular superdiffusion of turbulent magnetic fields (\cite{lazarian2014superdiffusion}), CRs do not experience trapping. Instead, they bounce among different magnetic mirrors and move diffusively along the local magnetic field, which leads to a new diffusion mechanism termed ``mirror diffusion". The mirror diffusion accounts for the suppressed diffusion of CRs when the mirroring condition is satisfied (\cite{xu2021mirror_supernova}; \cite{Barreto2023mirror}).

In this work, we will carry out the first numerical test of the mirror diffusion with test particle simulations in MHD turbulence. We will also examine the pitch-angle-dependent mean free path of mirror diffusion that is analytically predicted by LX21. The outline of this letter is as follows. In Section \ref{numeric}, we describe the numerical methods used for the test particle simulations. In Section \ref{results}, we present our numerical results on testing the mirror diffusion and measuring its mean free path in different MHD turbulence regimes. Our conclusions can be found in Section \ref{conclusions}.


\section{Numerical Methods}
\label{numeric}
For the turbulent magnetic fields used for our test particle simulations, we take snapshots from 3D compressible and incompressible MHD turbulence simulations after the MHD turbulence reaches a statistically steady state. The compressible MHD turbulence simulations are taken from \citet{yue2022superdiffusion}, which are generated via ZEUS-MP/3D code (\cite{hayes2006simulatingMHD}). It solves the ideal MHD equations in a periodic box under the isothermal condition. The turbulence is solenoidally driven in Fourier space with the injected turbulent kinetic energy peaked at wavenumber $k=2$, which is half of the data cube size, approximately 400 grids for the subsonic data cube (M1 henceforth) and 250 grids for the supersonic data cube (M2 henceforth). The numerical dissipation occurs at $l_{\text{diss}} \approx 20$ grids for M1 with numerical resolution $=792^3$ and $l_{\text{diss}} \approx 25$ grids for M2 with numerical resolution $=512^3$.  The incompressible MHD turbulence simulation (M3 henceforth) is taken from \cite{cho2010nonlocal}, which is generated by using the pseudo-spectral code developed by \cite{cho2000anisotropy}. It solves the incompressible MHD equations in a periodic box with hyperviscosity and hyperdiffusivity. The peak of turbulent energy injection is at $k\approx 2.5$ (corresponding to the turbulence injection scale $L_\text{inj} \approx 200$ grids), and the numerical resolution is $512^3$. The numerical dissipation occurs around $l_{\text{diss}} \approx 10$ grids. The inertial range of MHD turbulence is within $[L_\text{inj}, l_{\text{diss}}]$. In M1 and M2, the mean magnetic field $\vec{B}_0$ is in the z direction and in M3 $\vec{B}_0$ is in the x direction. By applying the hyperviscosity and hyperdiffusivity, M3 has a slightly more extended inertial range compared to M2 with the same resolution. Further details about these simulations can be found in \cite{yue2022superdiffusion}, \cite{cho2000anisotropy} and \cite{cho2010nonlocal}.
Since we are interested in the diffusion of relativistic CRs in non-relativistic MHD turbulence, that is, the test particle speed is much greater than the turbulent speed and the Alfv\'en speed $V_A$, the turbulent magnetic fields can be treated as a static background for the test particle and the motional electric field can be neglected 
\footnote{
We note that the properties of magnetic fields in a snapshot of MHD turbulence simulations are fundamentally different from static magnetic fields. The information on the stochasticity and superdiffusion of turbulent magnetic fields is well preserved in the snapshot. }. The parameters of the three MHD simulations used in this work are summarized in Table \ref{data}. 
MHD turbulence is characterized by the Alfv\'en Mach number $M_A = V_L / V_A \approx \langle v \rangle / \left(\langle B \rangle /\sqrt{4\pi \langle \rho \rangle}\right)$,  
where $V_L$ is the turbulent velocity at $L_\text{inj}$, $\langle v \rangle$, $ \langle\rho\rangle$ and $\langle B \rangle$ are the rms values of velocities, densities, and magnetic field strength. \footnote{Note that $M_A$ is defined as $V_L/V_A$ (\cite{lazarian2006enhancement}).
In practice, different methods for measuring $M_A$ in MHD turbulence simulations are introduced (e.g., \cite{xuyan2013}; \cite{cohet2016cosmic}; \cite{beattie2022ma}).} 
The three MHD simulations we use are sub-Alfv\'enic with $M_A < 1$.
\footnote{ {In super-Alfv\'enic turbulence with $M_A>1$, field line wandering becomes important in affecting the CR diffusion (\cite{brunetti2014cosmic}). Its comparison with mirror diffusion will be carried out in our future work. }} 
For compressible MHD turbulence, $M_s = V_L / c_s$ is the sonic Mach number, where $c_s$ is the isothermal sound speed. We include both subsonic ($M_s<1$) 
and supersonic ($M_s>1$) 
cases for the compressible MHD turbulence.

For our test particle simulations, we inject relativistic test particles that represent CR particles in a snapshot of MHD turbulence simulations. To trace the test particle trajectory, we numerically solve the following equations:  
\begin{equation}
    \frac{d\Vec{u}}{dt} = \frac{q}{\gamma mc}\Vec{u}\times \vec{B}~,\label{lorentz}
\end{equation}
and
\begin{equation}
    \frac{d\Vec{r}}{dt} = \Vec{u}~,\label{velocity}
\end{equation}
for a particle of position $\Vec{r}$, velocity $\Vec{u}$, charge $q$ and mass $m$. $\vec{B}$ is the magnetic field, $c$ is the speed of light, and $t$ is the time.
We follow a numerical method similar to that adopted in \cite{xuyan2013}; \cite{yue2022superdiffusion}. We use the Bulirsch-Stoer method to integrate Eq. \ref{lorentz} along with \ref{velocity} with an adaptive time step (\cite{press1988numerical}). It provides a high accuracy for the particle trajectory integration with a relatively low computational effort (see also \cite{lopez2016cosmic}).  
At each time step, the local magnetic field at the location of a test particle is interpolated by the cubic spline using the magnetic fields of the $10^3$ neighbouring points. 

The energy of a test particle is well conserved during the simulation and is quantified by the ratio $r_L/L_\text{inj}$, where the Larmor radius $r_L$ is defined as 
\begin{equation}
    r_L = \gamma\frac{mc^2}{q B_0}~,
\end{equation}
with the Lorentz factor $\gamma$. 
{Throughout this paper, we use $L_\text{inj}$ to normalize length scales as $L_\text{inj}$ is relatively well determined in our simulations.}
The gyrofrequency is defined as $\Omega = u/r_L$. At every time step, we measure $ \mu = \cos \theta$, where $\theta$ is the angle between $ \vec{u}(\vec{r})$  and $\vec{B}(\vec{r})$ , i.e., the pitch angle of a test particle. To achieve statistically reliable results, we inject a sufficiently large number of test particles and take an ensemble average of the numerical measurements over all test particles.

\section{Numerical Results}
\label{results}

The mirror diffusion of CRs in MHD turbulence arises from both the mirror reflection by magnetic compressions and the perpendicular superdiffusion caused by the Alfv\'enic turbulence. The latter plays an essential role in enabling the parallel diffusion of mirroring particles (LX21), which would otherwise be trapped between two fixed magnetic mirror points (CK73). To numerically test the mirror diffusion, we will first demonstrate the perpendicular superdiffusion of CRs, which has been analytically studied (\cite{lazarian2014superdiffusion}; \cite{lazarian2023cosmic}). The previous numerical testing of perpendicular superdiffusion of CRs was performed at small $\theta$ (e.g., \cite{xuyan2013}; \cite{yue2022superdiffusion}). To demonstrate the perpendicular superdiffusion of mirroring CRs with large $\theta$ (see Section \ref{mirror_diffusion}), we will adopt large initial $\theta$.

\subsection{Perpendicular Superdiffusion of mirroring CRs}
\label{superdiffusion}
The perpendicular superdiffusion of CRs is caused by that of turbulent magnetic fields and takes place within the inertial range of MHD turbulence (\cite{lazarian2014superdiffusion}). To have initially small separations between the particle trajectories, 
we inject 50 beams of test particles, and each beam consists of 20 particles. Within each beam, the 20 particles are initially randomly distributed within a volume of size of around 5 to 8 grids, so that the average initial separation between a pair of particles is around 3 to 5 grids.  All test particles have the same initial pitch angle cosine $\mu_0\approx0.10$ and $r_L=0.01L_\text{inj}$. Within each beam, we measure the perpendicular separation $\delta \Tilde{x}$ of every pair of particle trajectories with respect to $\vec{B}_0$ and obtain its rms value for all pairs. We then take the average over all beams as the measured perpendicualr displacement $\left\langle\sqrt{\langle \delta \Tilde{x} ^2 \rangle}\right\rangle$.

\begin{figure}[H]
\gridline{\fig{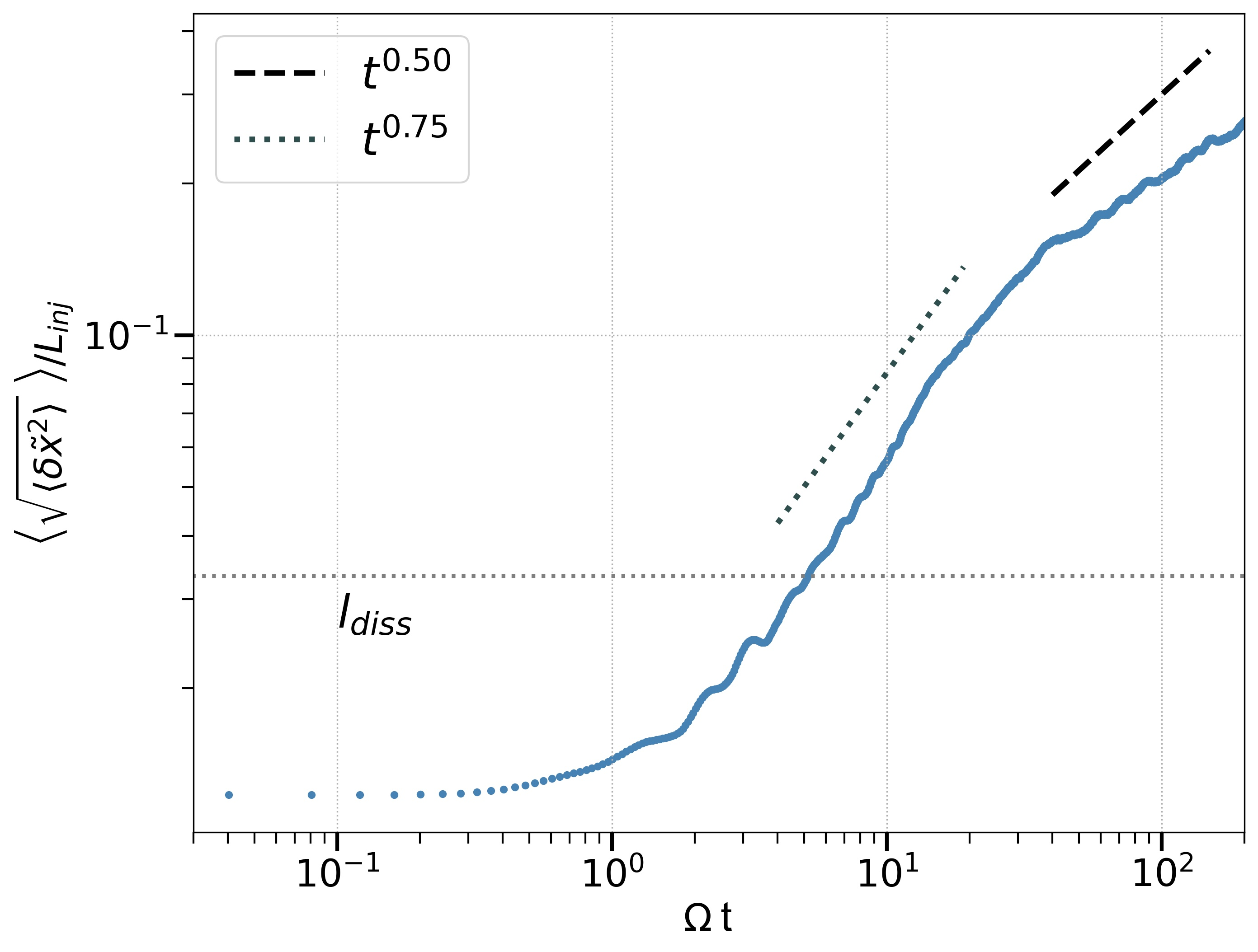}{0.44\textwidth}{(a) M1}}
\gridline{\fig{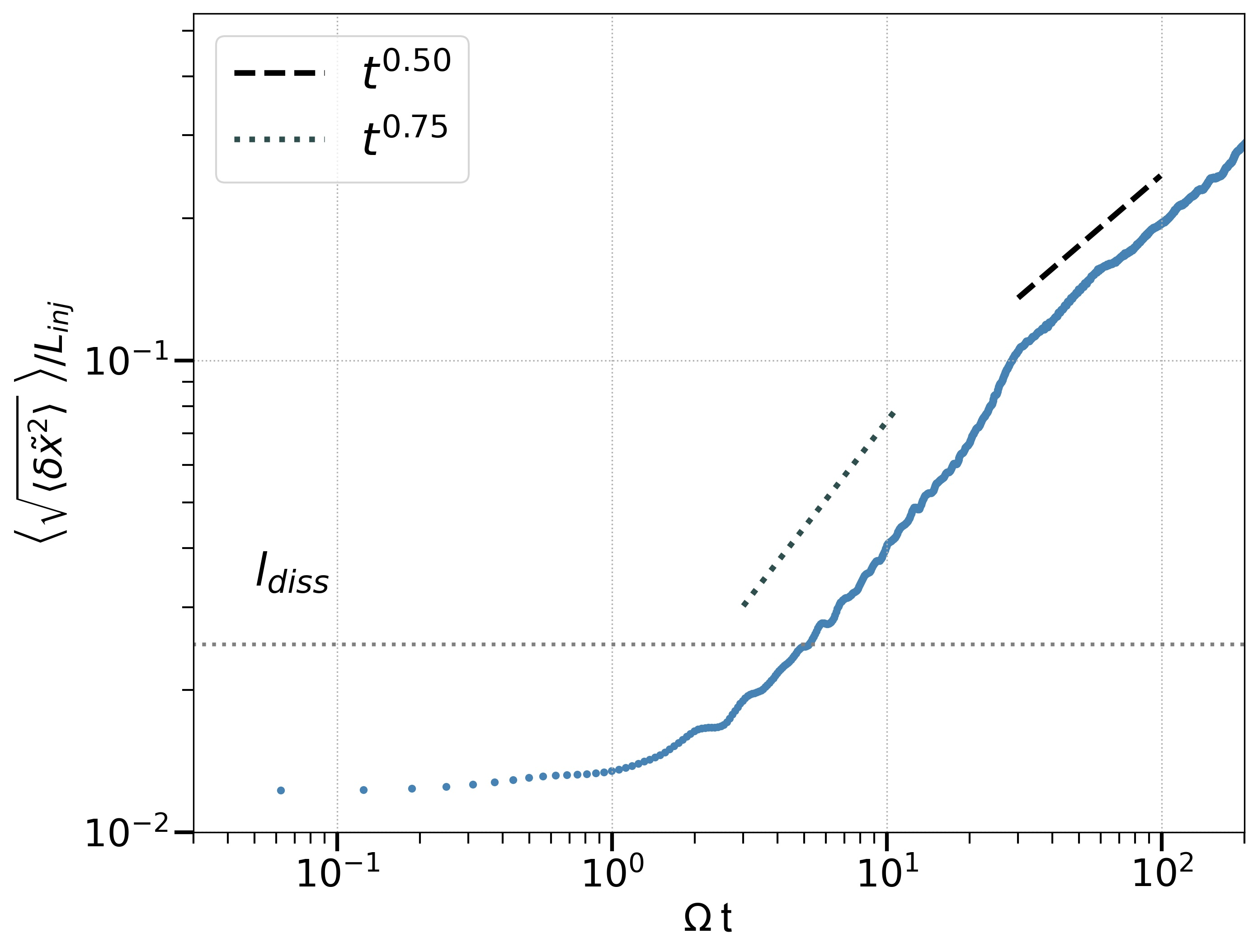}{0.44\textwidth}{(b) M2}}
\gridline{\fig{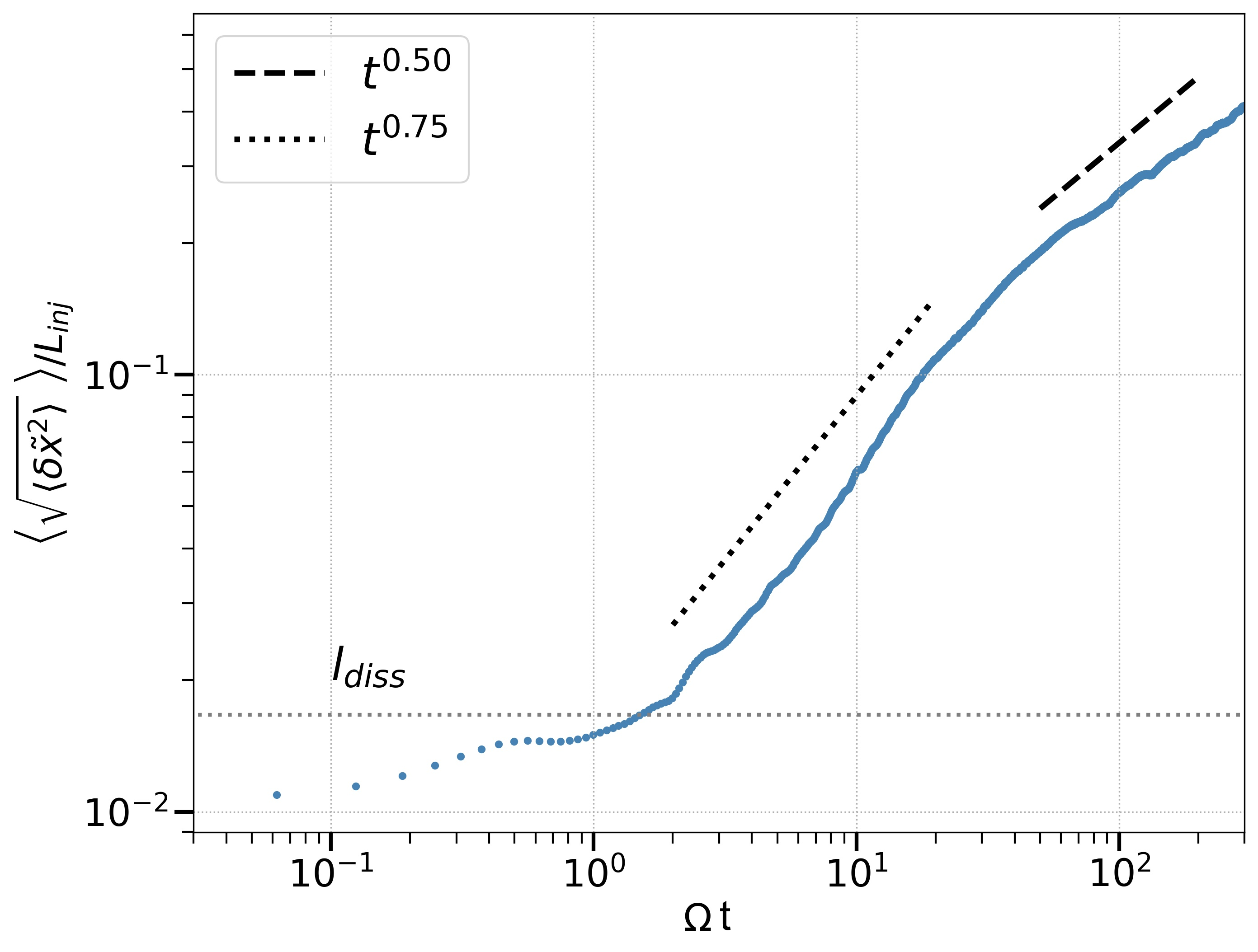}{0.44\textwidth}{(c) M3}}
\caption{{Averaged perpendicular separation of test particles} $\left\langle\sqrt{\langle \delta \Tilde{x} ^2 \rangle}\right\rangle$ vs. time for M1 (a), M2 (b) and M3 (c), with $r_L=0.01L_\text{inj}$ and {initial pitch angle cosine} $\mu_0=0.10$. $l_\text{diss}$ is indicated by the horizontal dashed line. }
\label{perp}
\end{figure}

The measured $\left\langle\sqrt{\langle \delta \Tilde{x} ^2 \rangle}\right\rangle$ normalized by $L_\text{inj}$ for M1, M2 and M3 as a function of $t$ (normalized by $\Omega^{-1}$) is presented in Fig. \ref{perp}. In all three cases, irrespective of the MHD turbulence regime, perpendicular superdiffusion is observed above $l_\text{diss}$ (see values in Table \ref{data}). Below $l_\text{diss}$, CRs undergo subdiffusion with $\sqrt{\langle \delta x ^2 \rangle}\propto t^{\alpha }$ and $\alpha<0.5$. 
Above $l_\text{diss}$, we see the transition from perpendicular superdiffusion with $\alpha>0.5$ to perpendicular normal diffusion with $\alpha \approx 0.5$. For the perpendicular superdiffusion, $\alpha =0.75$ is expected for the case when CRs undergo diffusive motion along the magnetic field (\cite{lazarian2023cosmic}). Our result indicates that the mirroring CRs simultaneously undergo parallel diffusion and perpendicular superdiffusion within the inertial range of MHD turbulence.

\begin{figure}[h]
\gridline{\fig{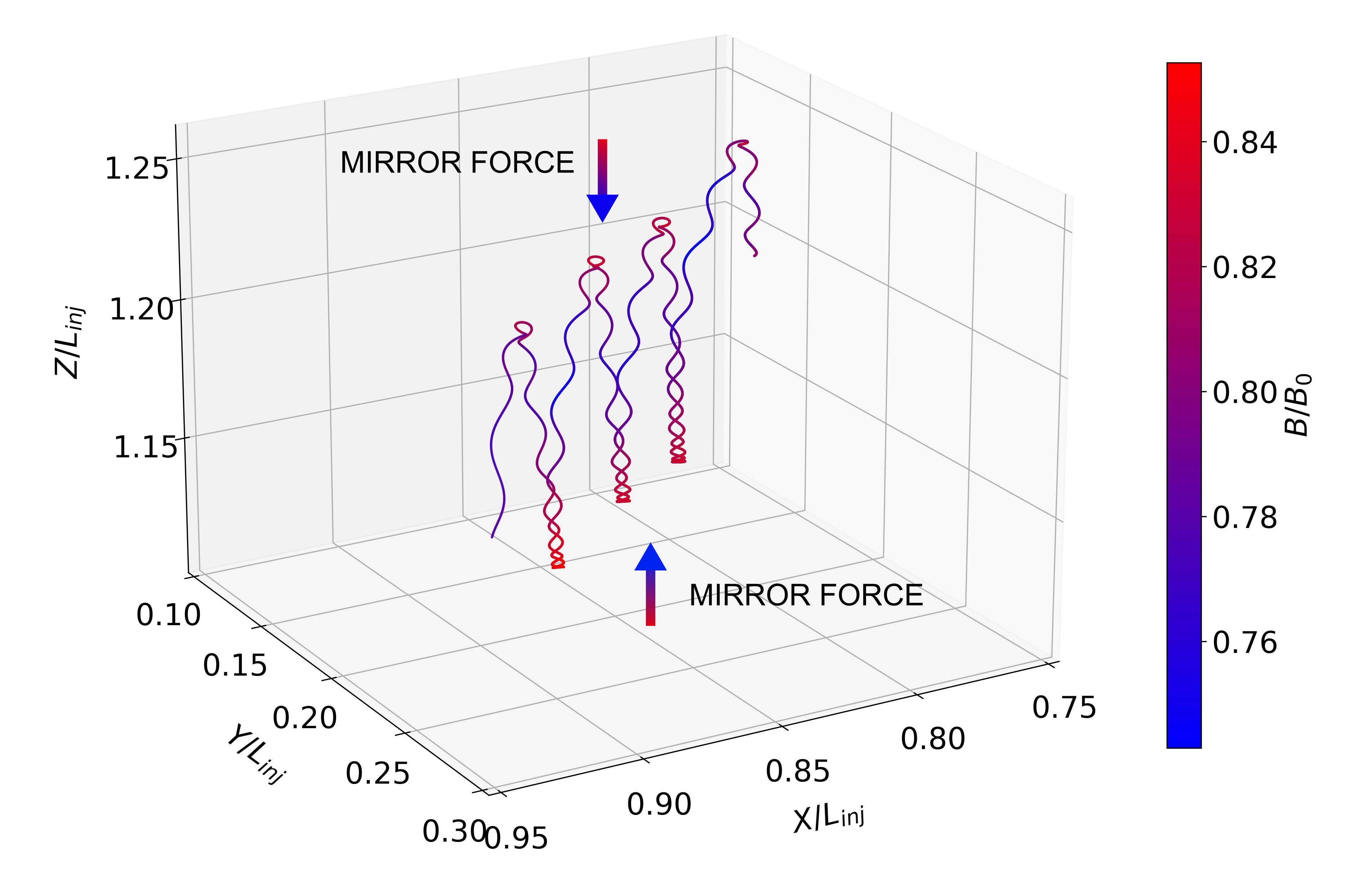}{0.53\textwidth}{(a)}}
\gridline{\fig{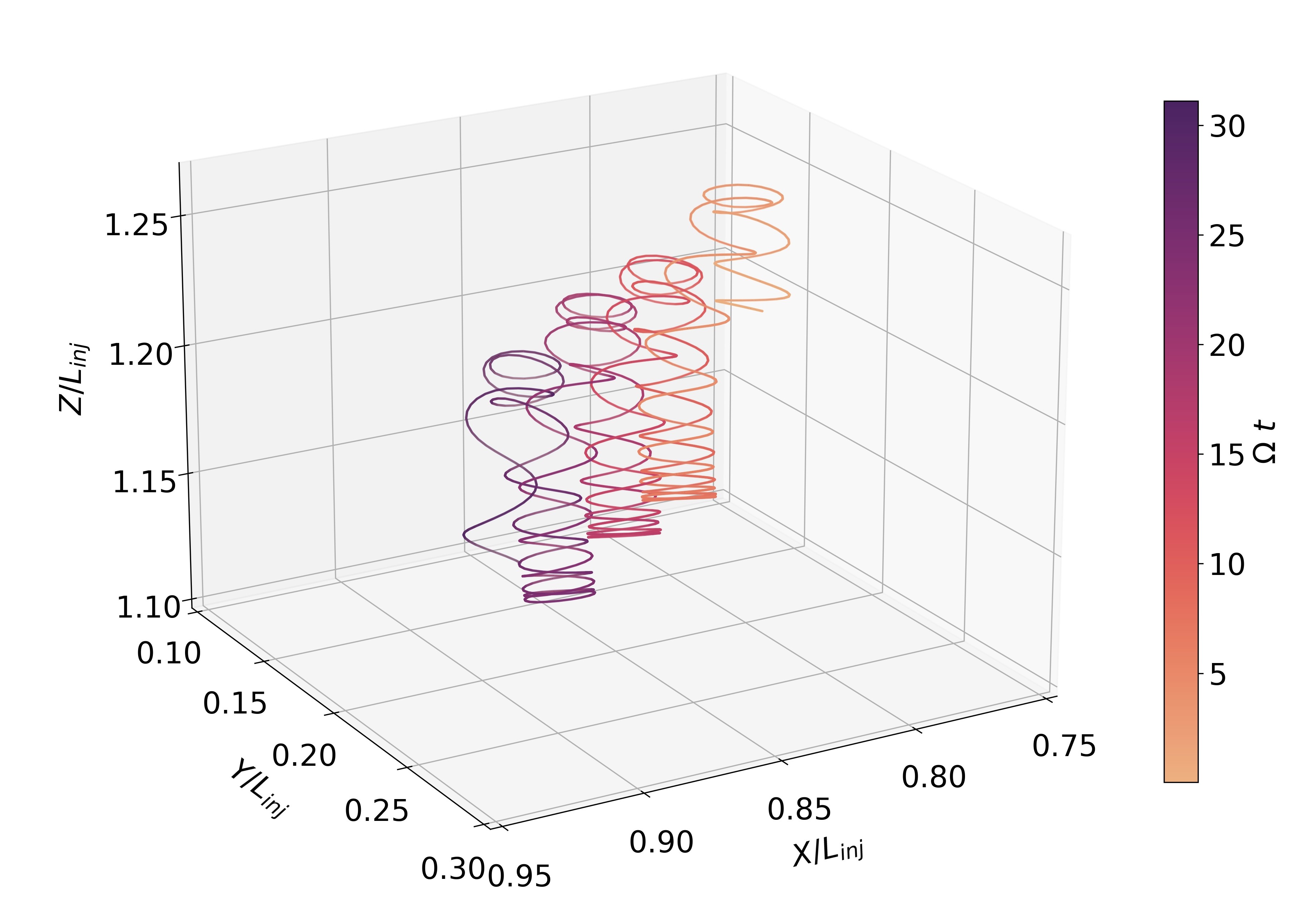}{0.53\textwidth}{(b)}}
\caption{Trajectory of a test particle in M2 with  $\mathbf{r_L = 0.01L_\text{inj}}$ and $\mu_0=0.05$. The trajectory is color coded by magnetic field strength in (a) and time in (b). Trajectories in (a) and (b) are the same except that 
the trajectory in (a) is averaged over the 
gyro-period $\Omega^{-1}$ to better illustrate the motion of the test particle. {Arrows in (a) indicate the directions of mirror forces.}}
\label{mirror_force}
\end{figure}

\subsection{Parallel Mirror Diffusion}
\subsubsection{Mirror Diffusion}
\label{mirror_diffusion}

In the presence of a magnetic mirror with a longitudinal magnetic gradient, CRs with $r_L$ smaller than the mirror size $l_\parallel$ and $\mu$ satisfying 
\begin{equation}
    \mu<\sqrt{\frac{\delta B_\parallel}{B_0 + \delta B_\parallel}}\label{mu}
\end{equation}
can be reflected by the magnetic mirror under the mirror force, where $\delta B_\parallel$ is the parallel magnetic fluctuation over $l_\parallel$ and $B_0$ is the mean magnetic field strength. The mirror force is (CK73)
\begin{equation}
    F_\parallel = -M \frac{\partial B_\parallel}{\partial l_\parallel}~,\label{force}
\end{equation}
where $M$ is the magnetic moment
\begin{equation}
    M= \frac{\gamma m u_\perp^2}{2B} ~,\label{adiabatic}
\end{equation}
where $u_\perp$ is the perpendicular component of $\vec{u}$. For mirroring CRs with their motion along the magnetic field dominated by magnetic mirroring, {there is no stochastic change of $\mu$}, and $M$ is a constant, known as the first adiabatic invariant.
As an example, in Fig. \ref{mirror_force} we present the trajectory of one test particle with  $r_L = 0.01L_\text{inj}$ and $\mu_0 = 0.05$ in M2.  We adopt a sufficiently small $r_L$ ($< L_\text{diss}$) {for mirroring to dominate over gyroresonant scattering} and a sufficiently small $\mu_0$ for the condition in Eq. \eqref{mu} to be satisfied. {The smallest $r_L$ is limited by our numerical resolution. In fact, $r_L =0.01L_\text{inj}$ is close to the smallest numerically resolvable scale, i.e., the grid size. }  In Fig. \ref{mirror_force} (a), the particle trajectory is color coded by $B/B_0$, where $B$ is the magnetic field strength. The corresponding $F_\|$ points toward a region with a weaker $B/B_0$. 
It can be seen that the particle is reflected at different mirror points (i.e., regions with stronger B) multiple times and move back and forth in the direction parallel to the magnetic field.
However, in the direction perpendicular to the $\vec{B}_0$ in the z direction, the particle is not trapped due to the perpendicular superdiffusion of turbulent magnetic fields (see Section \ref{superdiffusion}). As expected, CRs are not trapped between two magnetic mirorr points, but diffusively move along the magnetic field in MHD turbulence due to the perpendicular superdiffusion. 

\begin{figure}[t]
    \centering
    \hspace*{-1.cm}\includegraphics[scale=0.30]{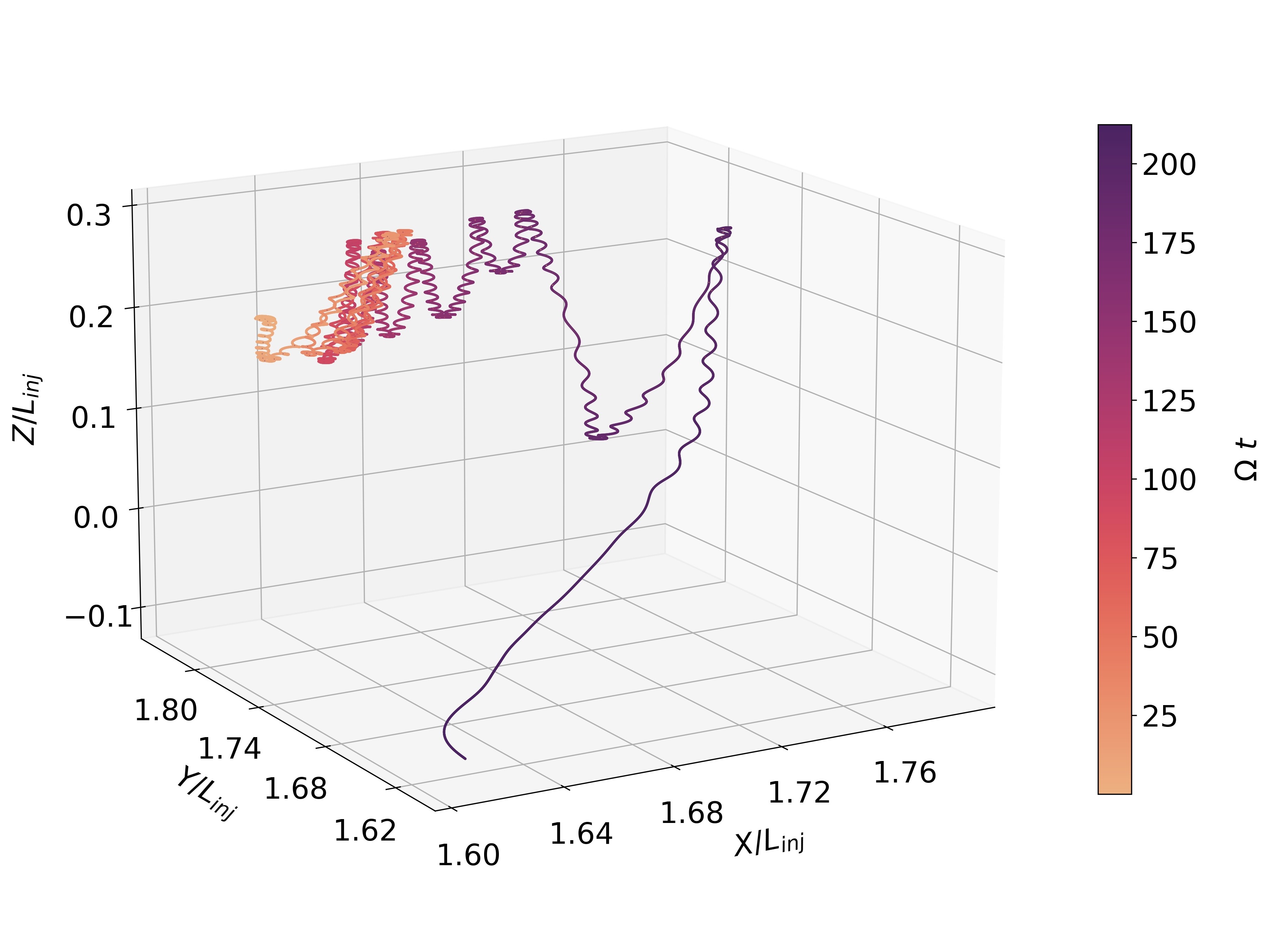}
    \caption{Gyro-averaged trajectory of a test particle in M2 with $r_L = 0.02L_\text{inj}$ and $\mu_0=0.05$. The trajectory is color coded by time, with a transition from mirror diffusion to scattering diffusion. }
    \label{mirror_scatter}
\end{figure}

\subsubsection{Transition between mirror diffusion and scattering diffusion}
\label{transition}

CRs with sufficiently small $\mu$ (see Eq. \ref{mu}) are subject to the nonresonant mirroring interaction and undergo the mirror diffusion. For CRs with larger $\mu$, their parallel diffusion is dominated by the resonant pitch-angle scattering and undergo the scattering diffusion (LX21). The transition between mirror diffusion and scattering diffusion takes place as $\mu$ stochastically changes due to the gyroresonant scattering. According to LX21, the transition between the two diffusion regimes is expected to occur at a critical $\mu$ as $\mu_c$. In compressible MHD turbulence, under the consideration that fast modes are mainly responsible for mirroring and scattering, $\mu_c$ is given by (LX21)

\begin{align}
    \mu_c = \min\{ \mu_\text{max}, \mu_\text{eq}\}~,\label{mu_c}
\end{align}
where 
\begin{equation}
    \mu_{\text{max}}=\sqrt{\frac{\delta B_f}{B_0 + \delta B_f}}~,\label{mu_max}
\end{equation}
and
\begin{equation}
    \mu_{\text{\text{eq}}}\approx\left[ \frac{14}{\pi} \left( \frac{\delta B_f}{B_0}\right)^{2} \left(\frac{r_L}{ L_\text{inj}}\right)^{\frac{1}{2}}\right]^{\frac{2}{11}} ~
\end{equation}
corresponds to the pitch angle cosine where mirroring and scattering are in balance, with $\delta B_f$ as the magnetic fluctuation at $L_\text{inj}$ of fast modes. Via mode decomposition (see \cite{cho2003compressible}; \cite{12statisticaldecomposition}), we find that $\delta B_f/B_0$ is close to $0.40$ for both M1 and M2, yielding $\mu_\text{max}\approx0.52$. For incompressible MHD turbulence, pseudo-Aflv\'en modes are responsible for mirroring. Under the consideration of inefficient scattering by Alfv\'en and pseudo-Alfv\'en modes for CRs with $r_L \ll L_\text{inj}$, we have (LX21)
\begin{equation}\label{eq:mumaxinc}
    \mu_{\text{max}}=\sqrt{\frac{\delta B_s}{B_0 + \delta B_s}}~,
\end{equation}
where $ \delta B_s$ is the magnetic fluctuation of pseudo-Alfv\'en modes at $L_\text{inj}$. As pseudo-Alf\'ven modes are slaved to Alf\'ven modes (\cite{beresnyak2019turbulencebook}), $\delta B_s/B_0$ can be approximately by $M_A$. Therefore, we have $\mu_c \approx \sqrt{M_A/(1+M_A)}$ for M3. At $\mu < \mu_c$, mirroring dominates scattering, and CRs are expected to undergo the mirror diffusion. At $\mu > \mu_c$, scattering becomes dominant, and CRs are expected to undergo the scattering diffusion.  

Next we numerically demonstrate the transition between the two different diffusion regimes. {Here we choose a slightly larger $r_L$ compared to that in Fig. \ref{mirror_force} for particles to undergo more scattering. } As shown in Fig. \ref{mirror_scatter}, from the trajectory of a test particle with $r_L=0.02L_\text{inj}$ and $\mu_0 = 0.05$ in M2, we can clearly see the transition from the slow mirror diffusion within a small region to the fast scattering diffusion over a large distance. During the mirror diffusion, the particle is confined in the direction of $\vec{B}_0$ (the z direction) with multiple mirror reflections. While during the scattering diffusion, the particle does not reverse its moving direction and rapidly moves along the magnetic field.

\begin{figure}[t]
    \centering
    \hspace*{-.8cm}\includegraphics[scale=0.4]{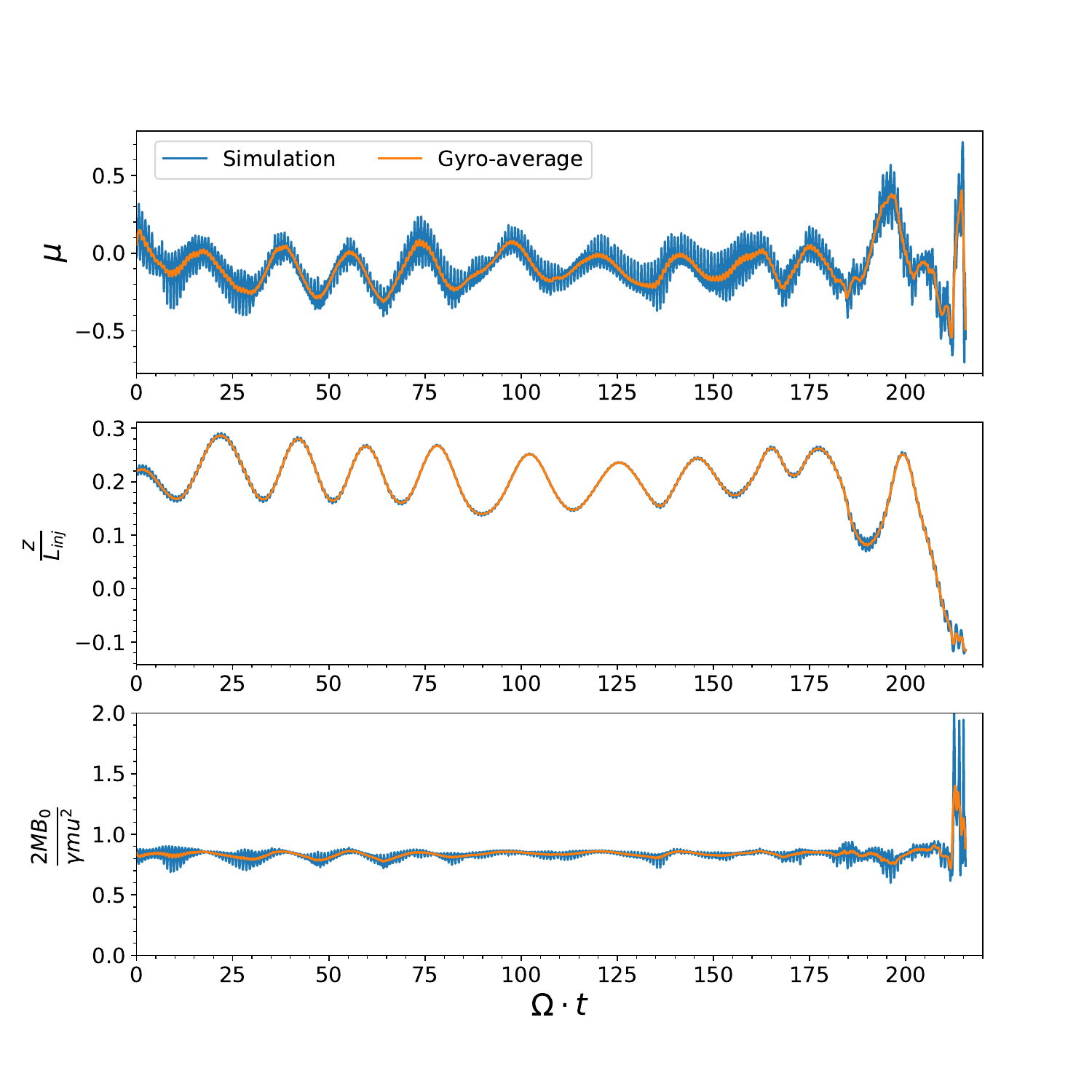}
    \caption{Time evolution of $\mu$, {particle position in z direction (the direction of the mean magnetic field for M2)} $z/L_\text{inj}$, and $2MB_0/\gamma mu^2$ corresponding to the test particle trajectory in Fig. \ref{mirror_scatter}. Blue and orange lines represent the numerical measurements and their gyro-averaged results, respectively.  }
    \label{xyz}
\end{figure}

Furthermore, in Fig. \ref{xyz} we present the time evolution of $\mu$, particle position in the z direction, and $2M/mu^2$ corresponding to the particle trajectory shown in Fig. \ref{mirror_scatter}. The numerical measurements and their gyro-averaged results are shown by blue and orange lines, respectively. Before $\Omega t \approx 200$, we see oscillations of $\mu$ around $\mu=0$, i.e., $\theta = 90^\circ$, which corresponds to the mirror reflection at the mirror points. Accordingly, the particle moves back and forth in the z direction (middle panel in Fig. \ref{xyz}).  After $\Omega t \approx 200$, $\mu$ changes stochastically, and the particle is not confined in the z direction. The transition between mirror diffusion and scattering diffusion occurs at $\mu\approx 0.5$, which is consistent with our estimate for $\mu_\text{max}$ in M2 (see Eq. \ref{mu_max}). 
As expected, M is conserved during the mirror diffusion, but varies dramatically during the scattering diffusion as efficient scattering violates the first adiabatic invariant (lower panel in Fig. \ref{xyz}). {The distinctive features of mirror diffusion and scattering diffusion can also been seen in simulations with a lower resolution 
(see Appendix).}

\begin{figure}[t]
  \centering
 
    \includegraphics[scale=0.45]{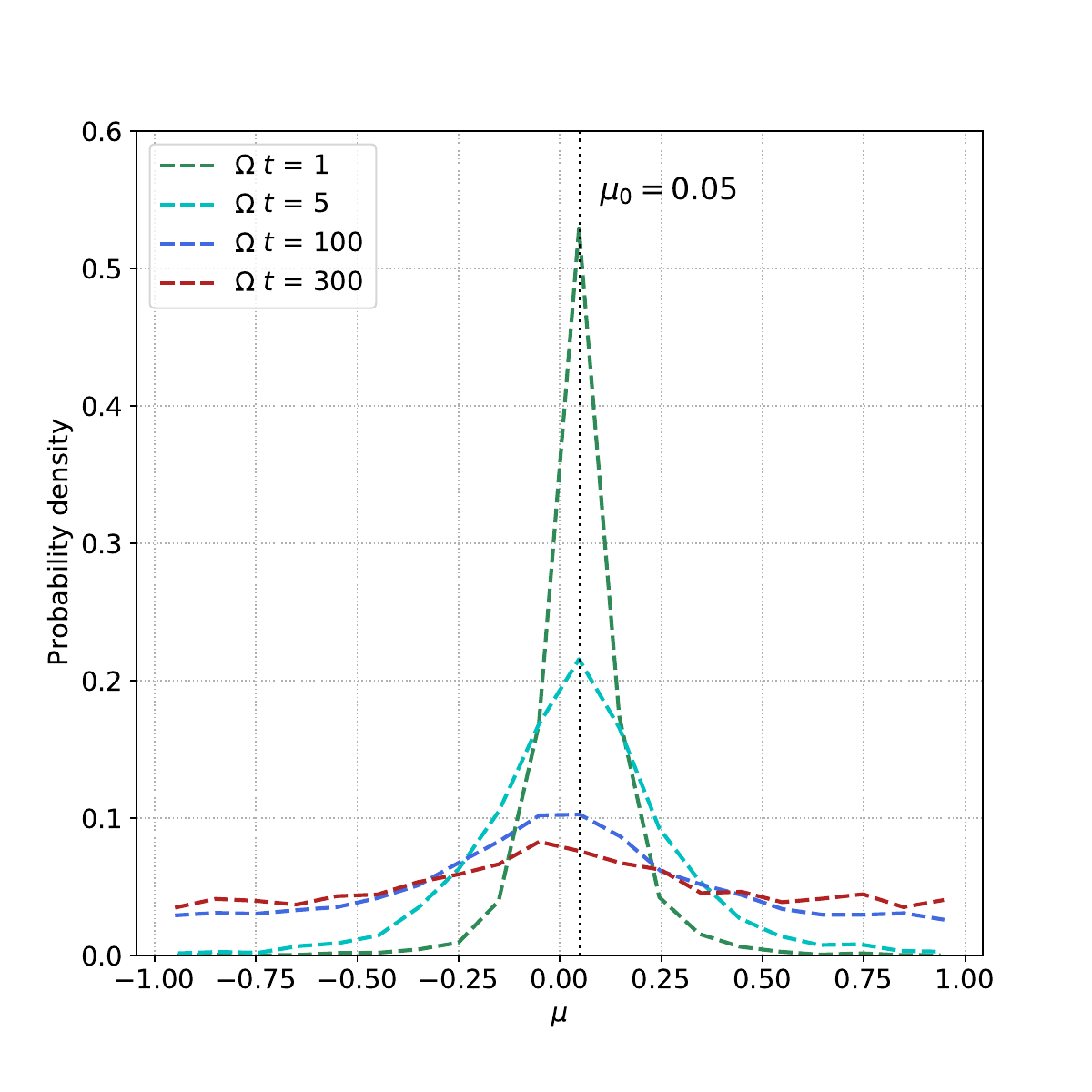}
  
    \caption{{Distributions of $\mu$ measured at different times for $5000$ test particles with $r_L = 0.01 L_\text{inj}$ and $\mu_0=0.05$ (indicated by the vertical dotted line) in M2. The bin size of $\mu$ is $0.05$. } }
    \label{mu_evo}
\end{figure}

The transition between mirror diffusion and scattering diffusion happens due to the stochastic change of $\mu$ caused by scattering. 
{To illustrate the stochastic change of $\mu$, in Fig. \ref{mu_evo}, we present the time evolution of the probability density distribution of $\mu$ of 5000 particles with $r_L = 0.01 L_\text{inj}$ and $\mu_0 = 0.05$ in M2. We see that due to the stochastic scattering, the distribution of $\mu$ gradually evolves into a more uniform distribution.}
As a result, CRs stochastically undergo the slow mirror diffusion when $\mu< \mu_c$ and the fast scattering diffusion when $\mu> \mu_c$, leading to a L\'{e}vy-flight-like propagation.

\begin{figure}[t]
\gridline{\fig{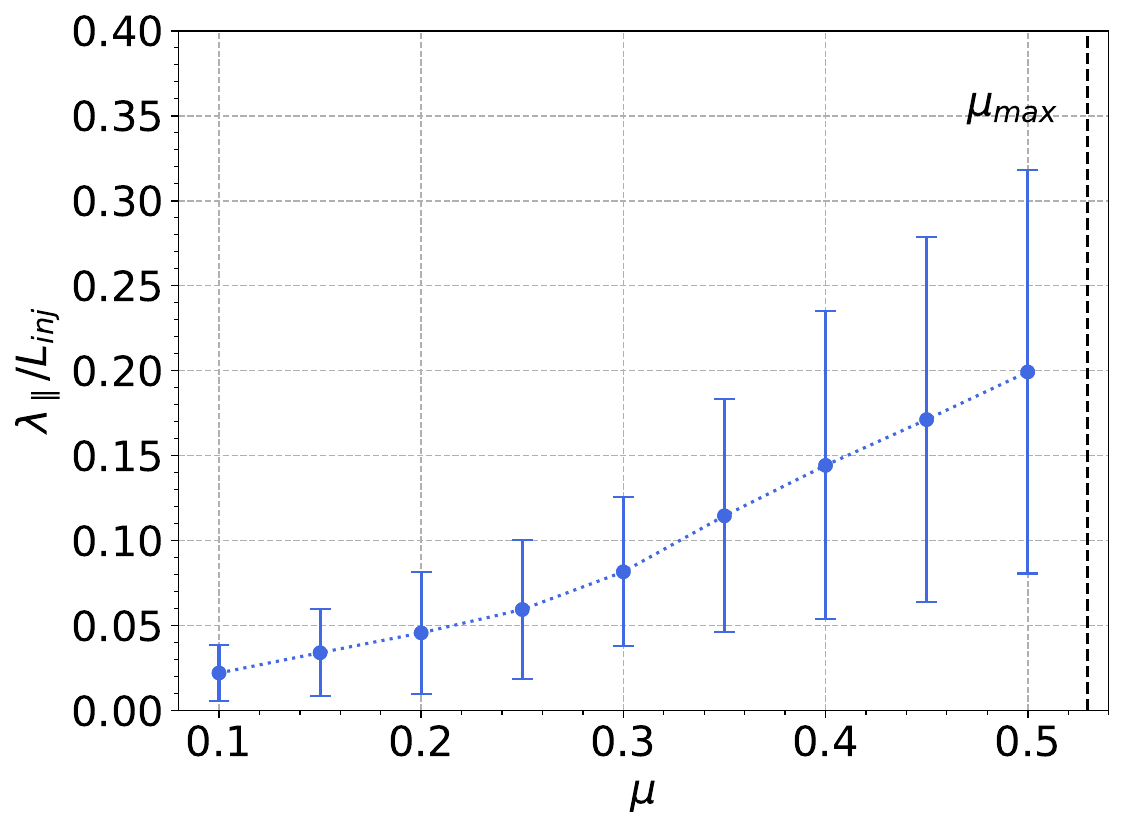}{0.43\textwidth}{(a) M1}}
\gridline{\fig{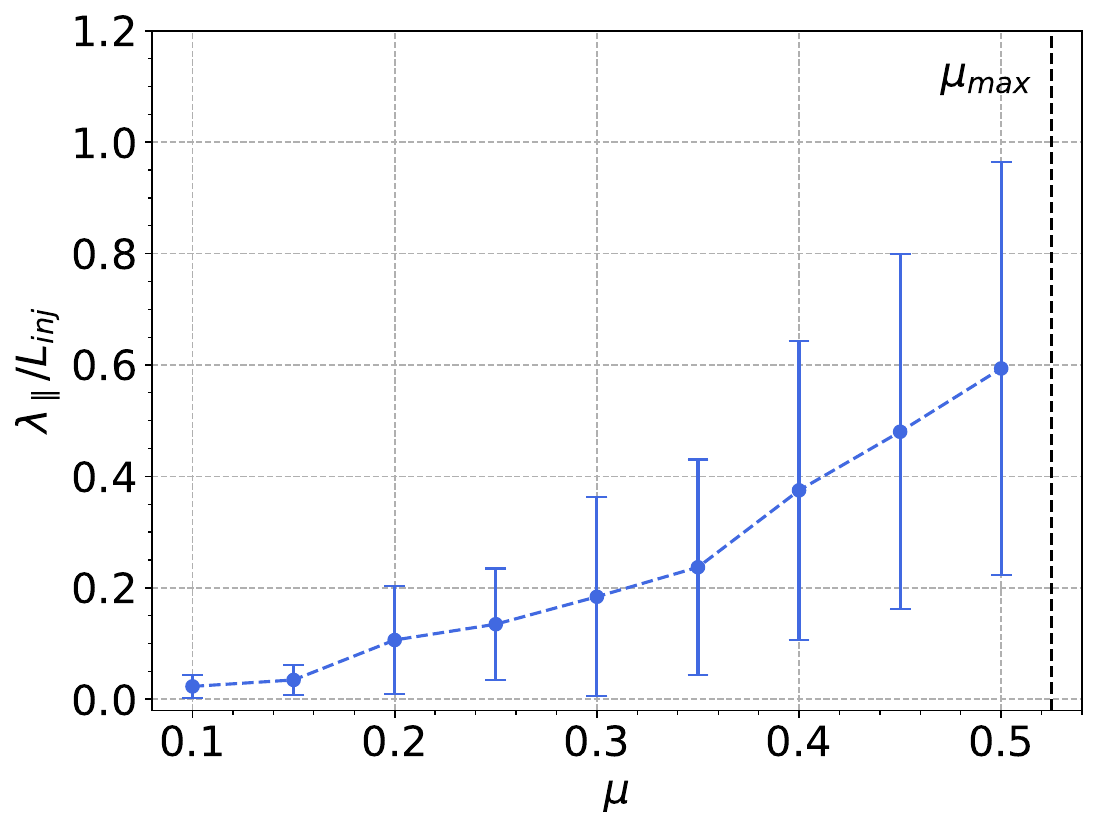}{0.43\textwidth}{(b) M2}}
\gridline{\fig{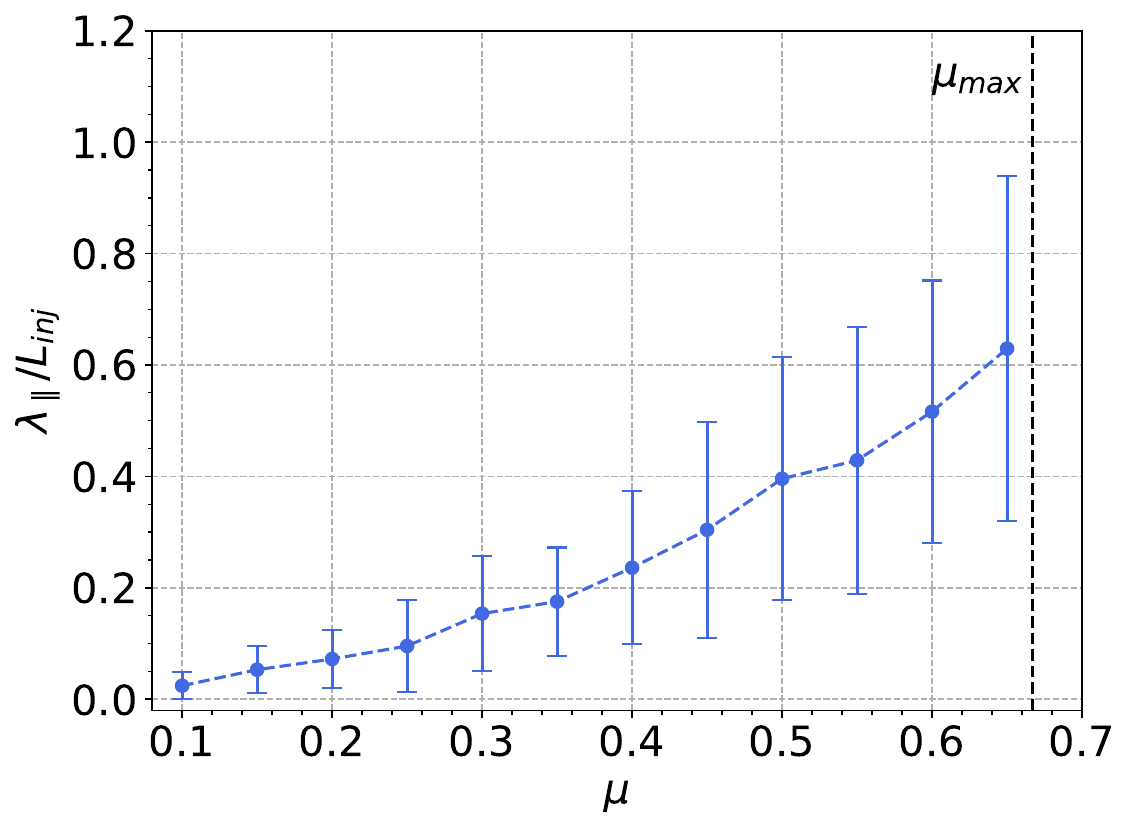}{0.43\textwidth}{(c) M3}}
\caption{Parallel mean free path $\lambda_\parallel/L_\text{inj}$ for mirror diffusion
measured at different $\mu$ for M1, M2, and M3. 
The vertical dashed lines indicate the estimated $\mu_\text{max}$, which is 
$\approx 0.52$ (Eq. \eqref{mu_max}) for M1 and M2 and $\approx 0.67$ (Eq. \eqref{eq:mumaxinc})
for M3. }
\label{mfp}
\end{figure}

\subsubsection{Pitch-angle-dependent Parallel Mean Free Path of Mirror Diffusion}
For the scattering diffusion, the scattering mean free path corresponds to the change of $\mu$ over the range $[\mu_c, 1]$ (CK73; LX21). In the presence of the mirror diffusion at $\mu<\mu_c$, the $90^\circ$ problem of scattering, i.e., the vanishing scattering near $90^\circ$ (\cite{90degreejokipii1966}), is naturally solved, and the scattering mean free path becomes finite. For the mirror diffusion, its parallel mean free path $\lambda_\|$ is determined by the size of the magnetic mirrors that the mirroring particles mainly interact with 
(LX21). For CRs with $\mu<\mu_c$, they preferentially interact with the mirrors satisfying (CK73; \cite{xu2020trapping})
\begin{equation}
    \mu = \sqrt{\frac{\delta b(l_\|)}{B_0 + \delta b(l_\|)}}~,
\end{equation}
where $l_\|$ is the mirror size, and $\delta b(l_\|)$ is the parallel magnetic fluctuation at $l_\|$. Along the energy cascade of MHD turbulence, toward smaller scales, $\delta b(l_\|)$ decreases, but the magnetic field gradient $\delta b(l_\|)/ l_\|$ increases, and the corresponding mirroring becomes more efficient (see Eq. \ref{force}). The above equation suggests that among the mirrors with sufficiently large magnetic fluctuations to reflect the CRs with $\mu$,  the smallest ones dominant the mirroring. Therefore,  $\lambda_\|$ of mirror diffusion is expected to depend on $\mu$ (LX21). {$\lambda_\parallel$ quantifies the slow diffusion of CR particles with sufficiently large pitch angles in the vicinity of CR sources (e.g., \cite{xu2021mirror_supernova}).}

To test the $\mu$-dependence of $\lambda_\|$, we inject test particles with different $\mu_0$. For the particles that exhibit oscillations within the same range $[-\mu, \mu]$, we measure their displacement $\Delta z$ in the z direction between two mirror points (see Fig. \ref{xyz} as an illustration) and take the average as the measured mean free path parallel to $\vec{B}_0$ corresponding to $\mu$. For sub-Alfv\'enic turbulence, magnetic field lines are mildly perturbed by turbulence especially at small scales. Our measurement serves as a good approximation of $\lambda_\|$ parallel to the local magnetic field at small $\mu$.

\begin{figure}[t]
  \centering
 
    \includegraphics[scale=0.33]{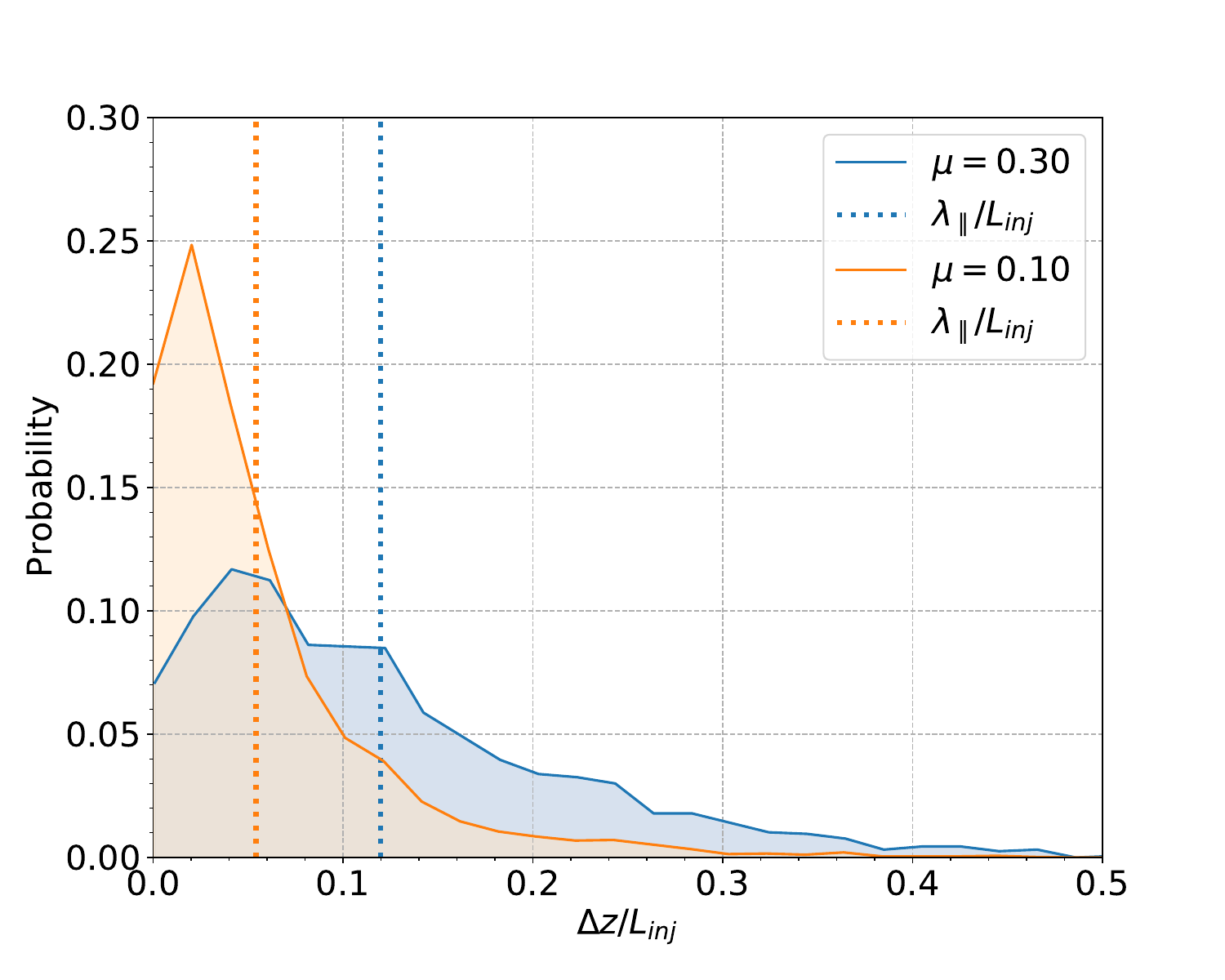}
  
    \caption{Distribution of parallel displacement $\Delta z$ measured for test particles with $r_L = 0.01 L_\text{inj}$ and different values of $\mu$ in M1. The bin size of $\Delta z$ is about $0.05L_\text{inj}$. The vertical dotted lines indicate the mean values.}
    \label{dis}
\end{figure}

In Fig. \ref{mfp}, we present the measured $\lambda_\|$ as a function of $\mu$ for M1, M2, and M3 {up to the theoretically estimated $\mu_\text{max}$ (Eqs. \eqref{mu_max} and \eqref{eq:mumaxinc}).} 
The width of each $\mu$ bin is 0.05, and the number of test particles in each $\mu$ bin ranges from 50 to 100 . We adopt $r_L=0.01L_\text{inj} \lesssim l_\text{diss}$  (see Table \ref{data}) for all the test particles to suppress the gyroresonant scattering, but the nonresonant mirroring is still active. Therefore, we have $\mu_c \approx \mu_\text{max}$ for both compressible and incompressible MHD turbulence.  
In all three MHD turbulence regimes, we see that $\lambda_\|$ increases with increasing $\mu$ within the range $\mu < \mu_\text{max}$, as predicted by LX21. The error bar represents the standard deviation $\sigma$ and also increases with increasing $\mu$. At a given $\mu$, we find that $\Delta z$ has a distribution (see Fig. \ref{dis}). As the degree of field line wandering increases toward larger scales, the deviation of the local magnetic field direction from the mean magnetic field direction becomes more significant at large scales. This may give rise to a larger $\sigma$ of the $\Delta z$ distribution at a larger $\mu$ (see Fig. \ref{dis}).

\section{conclusions}
\label{conclusions}
With test particle simulations in MHD turbulence, we performed the first numerical test of  mirror diffusion of CRs. Here we summarize our main findings.

(1) We identify the CRs with $\mu < \mu_c$ as mirroring CRs, which are subject to the mirror reflection by magnetic compressions in both compressible and incompressible MHD turbulence. We find that the mirroring CRs simultaneously undergo the superdiffusion perpendicular to the mean magnetic field and mirror diffusion along the turbulent magnetic field. Rather than being trapped between two fixed magnetic mirror points, the perpendicular superdiffusion enables encounters of mirroring CRs with different magnetic mirrors and their diffusion parallel to the magnetic field. 

(2) We numerically demonstrated the distinctive features of mirror diffusion and scattering diffusion. The diffusion of mirroring CRs is characterized by the oscillations in $\mu$ around $\mu=0$. As mirroring CRs repeatedly reverse their moving direction along the magnetic field, the mirror diffusion strongly enhances the confinement of CRs. By contrast, the scattering diffusion violates the first adiabatic invariant and is characterized by the stochastic change of $\mu$ within the range $[\mu_c, 1]$. As scattering CRs do not reverse their moving direction, they are weakly confined and rapidly diffuse along the magnetic field. The transition between the two diffusion regimes occurs due to the pitch-angle scattering. 

(3)  We numerically measured the parallel mean free path of mirror diffusion, which can be much smaller than the turbulence injection scale and increases wtih increasing $\mu$.

Our numerical measurements are in general consistent with the analytical predictions by LX21. As a new diffusion mechanism of CRs, the mirror diffusion has important implications. {Around CR sources, e.g., supernova remnants (\cite{xu2021mirror_supernova}), pulsar wind nebulae, the mirror diffusion of CRs with sufficiently large pitch angles} can account for their slow diffusion, as indicated by gamma-ray observations (e.g., \cite{abeysekara2017hwc}; \cite{evoli2018cosmic}). In addition, the mirror diffusion of pickup ions is important for interpreting the Interstellar Boundary Explorer ({\it IBEX}) ribbon (\cite{xuli2023}). {In the Galactic diffuse medium away from CR sources,} CRs stochastically undergo slow mirror diffusion and fast scattering diffusion. This L\'evy-flight-like propagation of CRs that we find is different from existing models of CR diffusion {solely} based on pitch-angle scattering. Its confrontation with CR observations will be further investigated.


\acknowledgments

We thank Yue Hu for helpful discussions. S.X. acknowledges the support for this work provided by NASA through the NASA Hubble Fellowship grant \# HST-HF2-51473.001-A awarded by the Space Telescope Science Institute, which is operated by the Association of Universities for Research in Astronomy, Incorporated, under NASA contract NAS5-26555.
This work used SDSC Expanse CPU in SDSC through the allocation PHY230032 from the Advanced Cyberinfrastructure Coordination Ecosystem: Services \& Support (ACCESS) program, which is supported by National Science Foundation grants \#2138259, \#2138286, \#2138307, \#2137603, and \#2138296.

\bibliographystyle{aasjournal}
\bibliography{bibli}

\appendix
\section{Mirror diffusion in a low-resolution MHD simulation}
{
Unlike the resonannt scattering, for the nonresonant mirroring interaction, the corresponding mirror diffusion of CRs at a given $\mu$ does not depend on $r_L$ as long as the length scale of magnetic fluctuations is larger than $r_L$. In this work, we focus on the mirror diffusion of CRs with $r_L \lesssim l_\text{diss}$, for which the resonant scattering is suppressed. This provides us a ``clean" test of mirror diffusion. 
For CRs with $r_L > l_\text{diss}$, the transition $\mu_c$ between mirror diffusion and scattering diffusion depends on $r_L$, and thus the overall diffusion also depends on $r_L$ (\cite{lazarian2021mirroring}). 
We will study the CR propagation with 
$r_L$ within the inertial range of turbulence in our future work.}

\begin{figure}[H]
  \centering
 
    \hspace*{-.8cm}\includegraphics[scale=0.4]{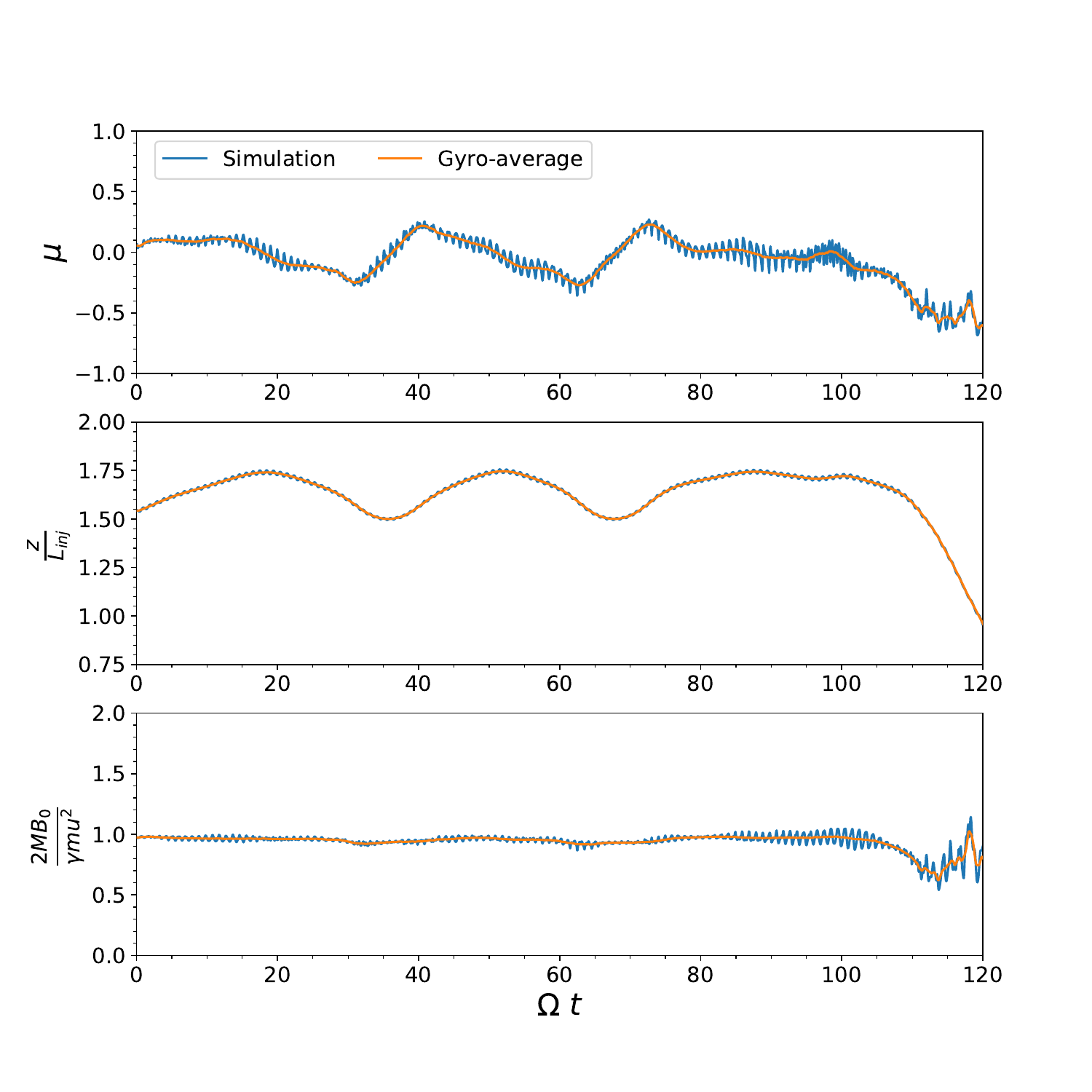}
    \caption{{Time evolution of $\mu$, {position in z direction} $z/L_\text{inj}$, and $2MB_0/\gamma mu^2$ for a test particle with $r_L=0.04L_\text{inj}$ and $\mu_0\approx 0.05$ in M4. Blue and orange lines represent the numerical measurements and their gyro-averaged results, respectively. } }
    \label{mirror_scatter_xyz_half}
\end{figure}

{
Here we examine whether the mirror diffusion of CRs with $r_L \lesssim l_\text{diss}$ can also be studied with a lower-resolution MHD simulation. 
We note that in a lower-resolution simulations, it becomes more challenging to both fully resolve $r_L$ and have $r_L$ much less than $l_\text{diss}$ given the more limited dissipation range and inertial range. 
Under these considerations, we adopt a slightly greater $r_L=0.04L_\text{inj}$ ($\lesssim l_\text{diss}$) for the test particle simulation in M4 of resolution $256^3$, which is obtained by smoothing M2 by averaging $8$ nearest neighboring grids. 
The parameters of M4 are listed in Table \ref{data_half}, 
which are similar to those of M2. 
To compare with the results obtained from M2, we measure the time evolution of $\mu$, $z$, and $M$ of test particles in M4 and show one example in Fig. \ref{mirror_scatter_xyz_half}. Similar to Fig. \ref{xyz}, this particle experiences mirroring with oscillations in $\mu$ and $z$ and constant $M$, but in this example, we see that the transition from mirrorring to scattering appears earlier, probably because the scattering 
is less suppressed 
as $r_L$ is closer to $l_\text{diss}$.}

\begin{table*}[htb!]
\centering
\caption{Parameters of MHD turbulence simulation with a low resolution\label{data_half}}
   
\begin{tabular}{|l||l l l l l l|} 
 \hline
   MHD Turbulence & $M_A$ & $M_s$ & $\beta$ & Resolution & $L_\text{inj}$ & $l_\text{diss}$ \\
 \hline\hline
 {Compressible supersonic (M4)}   & 0.41 & 10.44 & 0.003 & $256^3$ & $\approx 120$ & $\approx$ 8\\
 \hline
\end{tabular}
\centering 
\end{table*}


\end{CJK*}
\end{document}